\documentclass[twocolumn, journal]{IEEEtran}
\IEEEoverridecommandlockouts

\usepackage{diagbox}
\usepackage{bm}
\usepackage{color}
\usepackage{graphicx}
\usepackage{amsmath}
\usepackage{amssymb}
\usepackage{algorithm}
\usepackage{algorithmic}
\usepackage{ulem}
\usepackage{lineno}
\usepackage{lettrine}
\usepackage{subfigure}
\usepackage{epstopdf}
\usepackage{stfloats}
\usepackage{color}
\usepackage{graphicx}
\usepackage{amsmath}
\usepackage{amssymb}
\usepackage{algorithm}
\usepackage{algorithmic}
\usepackage{amsmath}
\usepackage{multirow}
\usepackage{booktabs}
\usepackage{array}
\usepackage{amsthm}
\usepackage{stfloats}
\usepackage{caption}
\usepackage{bm}
\usepackage{booktabs}
\usepackage{setspace}
\usepackage{makecell}
\usepackage{cases}

\newcommand{\eqdef}{\triangleq}
\newcommand{\be}{\begin{equation}}
\newcommand{\ee}{\end{equation}}
\newcommand{\bea}{\begin{eqnarray}}
\newcommand{\eea}{\end{eqnarray}}
\newcommand{\ba}{\begin{array}}
\newcommand{\ea}{\end{array}}

\title{\huge{Joint Transceiver Beamforming and Reflecting Design for Active RIS-Aided ISAC Systems}
}

\author{ \IEEEauthorblockN{Qi Zhu, Ming Li, \textit{Senior Member, IEEE}, Rang Liu, \textit{Graduate Student Member, IEEE},\\ and Qian Liu, \textit{Member, IEEE}}}

\begin{document}
\maketitle
\thispagestyle{empty}
\begin{abstract}
  Integrated sensing and communication (ISAC) is recognized as a promising technology with great potential in saving hardware and spectrum resources, since it simultaneously realizes radar detection and user communication functions in the fully-shared platform. Employing reconfigurable intelligent surface (RIS) in ISAC systems is able to provide a virtual line-of-sight (LoS) path to conquer blockage problem as well as introduce new degrees of freedom (DoFs) to further enhance system performance. Nevertheless, the multiplicative fading effect of passive RIS limits its applications in the absence of direct links, which promotes the development of active RIS.
  In this paper, we consider an active RIS-assisted ISAC system and aim to jointly design the transmit beamformer, the active RIS reflection and the radar receive filter to maximize the radar output signal-to-noise ratio (SNR) while guaranteeing pre-defined signal-to-interference-plus-noise ratios (SINRs) for communication users. To solve for this non-convex problem, an efficient algorithm is developed by leveraging the techniques of block coordinate descent (BCD), Dinkelbach's transform and majorization-minimization (MM). Simulation results verify the significant advancement of deploying active RIS in ISAC systems, which can achieve up to 32dB radar SNR enhancement compared with the passive RIS-assisted ISAC systems.
\end{abstract}


\section{Introduction}
Integrated sensing and communication (ISAC) has been envisioned as a promising enabler for future sixth-generation (6G) networks to provide high-quality wireless connectivity as well as highly accurate and robust sensing capability \cite{LiuF2}-\cite{LiuX}.
As another potential key technology of 6G, recently emerged reconfigurable intelligent surface (RIS) is popular owing to its superior ability to intelligently reconfigure wireless communication environment \cite{Wu}. Attracted by its significant advantages, RIS has been widely investigated in various communication scenarios \cite{Wu1}, \cite{ChenY}.

The applications of RIS in wireless communication motivate researchers to investigate the integration of RIS and ISAC systems to combine the advancements of both technologies \cite{Rang0}. Particularly, the deployment of RIS in ISAC systems can provide a virtual line-of-sight (LoS) path to conquer the blockage problem and introduce additional degrees of freedom (DoFs) to improve system performance.
The authors of \cite{WangX} employ an RIS in ISAC systems to mitigate multi-user interference (MUI) while ensuring the radar sensing beampattern.
The authors of \cite{LiuR} present comprehensive signal models for RIS-assisted ISAC systems and jointly design the transmit waveform, the receive filter, and the passive beamforming. The authors of \cite{LiuR1} further study sum-rate maximization of the communication system while satisfying the worst-case radar output signal-to-noise ratio (SNR).

While the advantages of RIS have been demonstrated, a plenty of studies on RIS have exposed its fatal problem, that is, multiplicative fading effect \cite{ZhangZ}.
In other words, the equivalent path loss of reflected source-RIS-destination link is usually several orders of magnitude larger than that of the direct source-destination link.
Due to this effect, conventional passive RIS cannot guarantee satisfactory performance when the direct link is unavailable or the users are not close enough to RIS.
Active RIS is recently emerging to overcome the multiplicative fading effect \cite{ZhangZ}, \cite{LongR}. Specifically, the novel active RIS still has the capacity to reflect the incoming signal with adjusted phase-shifts, and furthermore, each electromagnetic element of it is equipped with a dedicated amplifier to amplify the weak incident signal. The superiority of active RIS over the passive one in communication systems has been demonstrated in \cite{ZhangZ}-\cite{ChenY2}. However, the potential advantages of deploying active RIS in ISAC systems have not been exploited yet.

Motivated by the above discussions, in this paper, we investigate an active RIS-assisted ISAC system for enhancing both radar sensing and communication functionalities. Aiming at maximizing the radar output SNR as well as satisfying the communication users' quality-of-service (QoS) requirements, we propose to jointly design the transmit beamformer, the active RIS reflection coefficients and the radar receive filter. In order to handle the resulting non-convex problem, we decompose the original problem into several sub-problems, which can be alternately optimized by block coordinate descent (BCD) framework. Afterwards, efficient algorithms based on Dinkelbach's transform and majorization-minimization (MM) method are developed to solve for these sub-problems. Simulation studies verify the effectiveness of the proposed algorithm and validate the advantage of deploying active RIS in ISAC systems.

\section{System Model and Problem Formulation}

\begin{figure}[t]
\centering
  \includegraphics[width = 2.4 in]{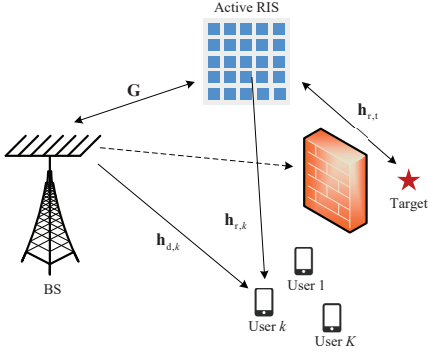}
  \caption{An active RIS-assisted ISAC system.}
  \label{fig:system model}
  \vspace{-0.5cm}
\end{figure}

Consider an active RIS-assisted ISAC system, in which a base station (BS) equipped with $N$ transmit/receive antennas\footnote{With advanced full-duplex technique, we assume that the transmit and receive antennas work simultaneously with perfect self-interference mitigation.} simultaneously serves $K$ communication users and detects a potential target blocked by obstacles\footnote{ Although the direct BS-target link is not considered in this work, the proposed algorithm can be easily extended to the direct-link-included cases.}, with the aid of an $M$-element active RIS.
As demonstrated in \cite{LiuX}, the transmit dual-functional signal is a weighted sum of communication symbols and radar signals, and given by
\begin{equation}\label{eq:transmit_signal}
  \mathbf{x} = \mathbf{W}_{\mathrm{c}}\mathbf{s}_{\mathrm{c}}+\mathbf{W}_{\mathrm{r}}\mathbf{s}_{\mathrm{r}} = \mathbf{W}\mathbf{s},
\end{equation}
where $\mathbf{s}_{\mathrm{c}} \in \mathbb{C}^K$ denotes the communication symbols intended for the $K$ users satisfying $\mathbb{E}\{\mathbf{s}_{\mathrm{c}} \mathbf{s}_{\mathrm{c}}^{H}\} =\mathbf{I}_K$ and $\mathbf{s}_{\mathrm{r}} \in \mathbb{C}^N$ denotes $N$ individual radar signals satisfying $\mathbb{E}\{\mathbf{s}_{\mathrm{r}} \mathbf{s}_{\mathrm{r}}^{H}\} =\mathbf{I}_N$. We assume that $\mathbf{s}_{\mathrm{c}}$ and $\mathbf{s}_{\mathrm{r}}$ are statistically independent and uncorrelated, i.e., $\mathbb{E}\{\mathbf{s}_{\mathrm{c}} \mathbf{s}_{\mathrm{r}}^{H}\} =\mathbf{0}$. $\mathbf{W}_{\mathrm{c}} \in \mathbb{C}^{N \times K}$ and $\mathbf{W}_{\mathrm{r}} \in \mathbb{C}^{N \times N}$ represent the corresponding beamforming matrices for the communication symbols and radar signals. Moreover, we define $\mathbf{W} \eqdef [\mathbf{W}_{\mathrm{c}}~\mathbf{W}_{\mathrm{r}}]$ and $\mathbf{s} \eqdef [\mathbf{s}_{\mathrm{c}}^T~\mathbf{s}_{\mathrm{r}}^T]^T$ for simplicity.

Denote $\mathbf{h}_{\mathrm{d}, k} \in \mathbb{C}^N$, $\mathbf{G}\in \mathbb{C}^{M \times N}$, and $\mathbf{h}_{\mathrm{r}, k} \in \mathbb{C}^M$ as the channels from the BS to the $k$-th user, from the BS to the RIS, and from the RIS to the $k$-th user, respectively\footnote{Since various efficient algorithms (e.g., \cite{AL}) have been proposed to handle the channel state information (CSI) acquisition problem for RIS-assisted systems, we assume perfect CSI in this paper.}.
Since an additional reflection-type amplifier is integrated to the electromagnetic element, the reflection coefficient of the $m$-th element of the active RIS can be denoted as $\phi_m \eqdef a_m e^{\jmath\varphi_m}$, where $a_m$ and $\varphi_m$ represent the amplitude and phase-shift, respectively. Considering the practical amplification ability of the amplifier, we assume the amplitude $a_m$ is constrained by the maximum amplification gain, i.e., $a_m \leq a_{\max},~\forall m$. Let $\mathbf{\Phi} \eqdef \text{diag}\{\boldsymbol{\phi}\}$ represent the reflection beamforming matrix of the active RIS with $\boldsymbol{\phi} \eqdef [ \phi_1,\cdots, \phi_M]^T \in \mathbb{C}^M$.
Then, the received signal at the $k$-th user can be expressed as
\begin{equation}\label{eq:received_signal_user}
  y_k = (\mathbf{h}_{\mathrm{d}, k}^T + \mathbf{h}_{\mathrm{r}, k}^T\mathbf{\Phi}\mathbf{G})\mathbf{x} + \mathbf{h}_{\mathrm{r}, k}^T\mathbf{\Phi}\mathbf{z}_0 + n_k,
\end{equation}
where $\mathbf{z}_0 \sim \mathcal{C} \mathcal{N} (\mathbf{0}_M, \sigma^{2}_\mathrm{z}\mathbf{I}_{M})$ and $n_k \sim \mathcal{C} \mathcal{N} (0, \sigma_k^{2})$ are the dynamic noise at the active RIS and the additive white Gaussian noise (AWGN) at the $k$-th user, respectively. Thus, the signal-to-interference-plus-noise ratio (SINR) of the $k$-th user can be calculated as
\vspace{-0.2cm}
\begin{equation}\label{eq:SINR_at_user_k}
  \gamma_{k}=\frac{|\mathbf{h}_{k}^{T} \mathbf{w}_{k}|^{2}}{\sum_{i \neq k}^{K+N}|\mathbf{h}_{k}^{T} \mathbf{w}_{i}|^{2}+\|\mathbf{h}_{\mathrm{r}, k}^{T} \mathbf{\Phi}\|^{2}_2 \sigma_{\mathrm{z}}^{2}+\sigma_k^{2}},
\end{equation}
where we define $\mathbf{h}_{k}^{T} \triangleq \mathbf{h}_{\mathrm{d}, k}^{T} + \mathbf{h}_{\mathrm{r}, k}^{T} \mathbf{\Phi} \mathbf{G}$ as the composite channel from the BS to the $k$-th user and $\mathbf{w}_k$ as the $k$-th column of $\mathbf{W}$, i.e., $\mathbf{W} \eqdef [\mathbf{w}_1,\cdots, \mathbf{w}_{K+N}]$.

Meanwhile, considering that the direct BS-target link is unavailable, the received echo signal through BS-RIS-target-RIS-BS path can be expressed as\vspace{0.05cm}
\begin{equation}\label{eq:received_signal_BS}
  \mathbf{y}_{\mathrm{r}} = \mathbf{G}^T\mathbf{\Phi}\left[\mathbf{h}_{\mathrm{r}, \mathrm{t}} \beta \mathbf{h}_{\mathrm{r}, \mathrm{t}}^T\mathbf{\Phi}(\mathbf{G}\mathbf{x} + \mathbf{z}_{0}) + \mathbf{z}_1 \right]+ \mathbf{n}_{\mathrm{r}},
\end{equation}
where the scalar $\beta$ represents the target radar cross section (RCS) with $\mathbb{E}\{|\beta|^2\} = \varsigma_{\mathrm{t}}^2$. $\mathbf{h}_{\mathrm{r}, \mathrm{t}} \in \mathbb{C}^M$ denotes the channel between the RIS and the target. It is worthy noting that the RIS-target link is LoS and the angle of arrival/departure (AoA/AoD) of interest is known a priori. $\mathbf{z}_1$ is the introduced dynamic noise of the active RIS when the radar signal returns, which is independent and identically distributed with $\mathbf{z}_0$. $\mathbf{n}_\mathrm{r} \sim \mathcal{C} \mathcal{N} (\mathbf{0}_N, \sigma_\mathrm{r}^{2}\mathbf{I}_{N})$ is AWGN. After processing the received signal $\mathbf{y}_{\mathrm{r}}$ with a receive filter $\mathbf{u} \in \mathbb{C}^{N}$, the radar output signal at the BS is written as
\begin{equation}\label{eq:signal_after_filter}
  \mathbf{u}^H\mathbf{y}_\mathrm{r} = \beta \mathbf{u}^H \mathbf{H}_\mathrm{t}\mathbf{x} + \beta \mathbf{u}^H \mathbf{H}_{\mathrm{z},0} \mathbf{z}_0 + \mathbf{u}^H \mathbf{H}_{\mathrm{z},1}\mathbf{z}_{1} + \mathbf{u}^H \mathbf{n}_{\mathrm{r}},
\end{equation}
where for notational simplicity, we define $\mathbf{H}_\mathrm{t}\eqdef \mathbf{G}^T\mathbf{\Phi}\mathbf{h}_{\mathrm{r}, \mathrm{t}}\mathbf{h}_{\mathrm{r}, \mathrm{t}}^T\mathbf{\Phi}\mathbf{G}$, $\mathbf{H}_{\mathrm{z},0} \eqdef \mathbf{G}^T\mathbf{\Phi}\mathbf{h}_{\mathrm{r}, \mathrm{t}}\mathbf{h}_{\mathrm{r}, \mathrm{t}}^T \mathbf{\Phi}$ and $\mathbf{H}_{\mathrm{z},1} \eqdef \mathbf{G}^T\mathbf{\Phi}$.
Thus, the radar output SNR is obtained as
\begin{equation}\label{eq:SINR_r}
\gamma_\mathrm{r} = \frac{\varsigma_\mathrm{t}^2 \mathbf{u}^H\mathbf{H}_\mathrm{t}\mathbf{W}\mathbf{W}^H\mathbf{H}_\mathrm{t}^H\mathbf{u}}
{\mathbf{u}^H(\varsigma_\mathrm{t}^2\sigma_\mathrm{z}^2\mathbf{H}_{\mathrm{z},0}\mathbf{H}_{\mathrm{z},0}^H + \sigma_\mathrm{z}^2\mathbf{H}_{\mathrm{z},1}\mathbf{H}_{\mathrm{z},1}^H + \sigma_\mathrm{r}^2\mathbf{I}_N)\mathbf{u}}.
\end{equation}

In this paper, we aim to jointly design the transmit beamforming $\mathbf{W}$, the active RIS reflection coefficients $\boldsymbol{\phi}$, and the radar receive filter $\mathbf{u}$ to maximize the radar SNR $\gamma_\mathrm{r}$, while satisfying the communication users' QoS requirements $\Gamma_k$, the power budgets $P_\mathrm{BS}$ at the BS and $P_\mathrm{RIS}$ at the active RIS, and the maximum amplitude $a_\mathrm{\max}$ of the reflecting coefficients. Therefore, the optimization problem is formulated as
\vspace{-0.1cm}
\begin{subequations}\label{pr:original_problem}
\begin{align}
\max_{\mathbf{W}, \boldsymbol{\phi}, \mathbf{u}} ~~&\gamma_\mathrm{r}\\
\text{s.t.} ~~~~& \|\mathbf{W}\|^2_{F} \leq P_{\mathrm{BS}},\\
& \mathcal{P}(\mathbf{W},\boldsymbol{\phi}) \leq P_{\mathrm{RIS}},\\
& \gamma_{k} \geq \Gamma_{k},~\forall k,\\
& a_m \leq a_{\max},~\forall m,
\end{align}
\end{subequations}
where
\begin{equation}\label{eq:P}
\begin{aligned}
  \mathcal{P}(\mathbf{W},\boldsymbol{\phi}) &= \|\mathbf{\Phi}\mathbf{G}\mathbf{W}\|^2_F + \varsigma_\mathrm{t}^2\|\mathbf{\Phi}\mathbf{h}_{\mathrm{r}, \mathrm{t}}\mathbf{h}_{\mathrm{r}, \mathrm{t}}^T\mathbf{\Phi}\mathbf{G}\mathbf{W}\|^2_F\\
   & \quad +\varsigma_\mathrm{t}^2\sigma_\mathrm{z}^2\|\mathbf{\Phi}\mathbf{h}_{\mathrm{r}, \mathrm{t}}\mathbf{h}_{\mathrm{r}, \mathrm{t}}^T\mathbf{\Phi}\|^2_{F} + 2\sigma_\mathrm{z}^2\|\mathbf{\Phi}\|^2_F,
\end{aligned}
\end{equation}
represents the reflection power consumed at the active RIS.
Clearly, problem (\ref{pr:original_problem}) is non-convex and highly challenging to solve due to the complicated and multi-variable coupling function (\ref{pr:original_problem}a), constraints (\ref{pr:original_problem}c) and (\ref{pr:original_problem}d). To tackle these difficulties, in the following, we propose to decompose it into several more tractable sub-problems and solve them in an alternating way.

\section{Joint Beamforming and Reflection Design}
\subsection{Receive Filter Design}
With given transmit beamforming $\mathbf{W}$ of the BS and reflection coefficients $\boldsymbol{\phi}$ of the active RIS, the optimization for the receive filter $\mathbf{u}$ can be expressed as
\begin{equation}\label{pr:u}
\max_{\mathbf{u}} \quad \frac{\varsigma_\mathrm{t}^2 \mathbf{u}^H\mathbf{A}\mathbf{u}}
{\mathbf{u}^H\mathbf{B}\mathbf{u}}
\end{equation}
where for brevity we define
\begin{subequations}\label{eq:trans_A_B}
\begin{align}
\mathbf{A} &\eqdef \mathbf{H}_\mathrm{t}\mathbf{W}\mathbf{W}^H\mathbf{H}_\mathrm{t}^H,\\
\mathbf{B} &\eqdef \varsigma_\mathrm{t}^2\sigma_\mathrm{z}^2\mathbf{H}_{\mathrm{z},0}\mathbf{H}_{\mathrm{z},0}^H + \sigma_\mathrm{z}^2\mathbf{H}_{\mathrm{z},1}\mathbf{H}_{\mathrm{z},1}^H + \sigma_\mathrm{r}^2\mathbf{I}_N.
\end{align}
\end{subequations}
It is obvious that problem (\ref{pr:u}) is a typical generalized Rayleigh quotient, whose optimal solution is the eigenvector corresponding to the largest eigenvalue of the matrix $\mathbf{B}^{-1}\mathbf{A}$.

\vspace{-0.2cm}
\subsection{Transmit Beamforming Design}
In this subsection, we focus our attention on optimizing the transmit beamforming $\mathbf{W}$, while fixing the receive filter $\mathbf{u}$ and the reflection coefficients $\boldsymbol{\phi}$. By defining
\begin{subequations}\label{eq:trans_W}
\allowdisplaybreaks[4]
\begin{align}
\mathbf{w} &\triangleq \text{vec}\{\mathbf{W}\}  = [\mathbf{w}_1^T, \cdots, \mathbf{w}_{K+N}^T]^T,\\
\mathbf{Y} &\triangleq \mathbf{I}_{K+N} \otimes \mathbf{H}_\mathrm{t}^H\mathbf{u}\mathbf{u}^H\mathbf{H}_\mathrm{t},\\
\mathbf{Z} &\triangleq \mathbf{I}_{K+N} \otimes (\mathbf{G}^H\mathbf{\Phi}^H\mathbf{\Phi}\mathbf{G} + \varsigma_\mathrm{t}^2 \widetilde{\mathbf{Z}}^H\widetilde{\mathbf{Z}}),\\
\widetilde{\mathbf{Z}} &\triangleq \mathbf{\Phi}\mathbf{h}_{\mathrm{r}, \mathrm{t}}\mathbf{h}_{\mathrm{r}, \mathrm{t}}^T\mathbf{\Phi}\mathbf{G},\\
{\mathbf{H}}_{k} &\eqdef \mathbf{I}_{K+N}\otimes\mathbf{h}_k^{T},\\
c_\mathrm{RIS} & \eqdef \varsigma_\mathrm{t}^2\sigma_\mathrm{z}^2\|\mathbf{\Phi}\mathbf{h}_{\mathrm{r}, \mathrm{t}}\mathbf{h}_{\mathrm{r}, \mathrm{t}}^T\mathbf{\Phi}\|^2_{F} + 2\sigma_\mathrm{z}^2\|\mathbf{\Phi}\|^2_F,\\
c_0 &\eqdef \|\mathbf{h}_{\mathrm{r}, k}^{T} \mathbf{\Phi}\|^{2}_2 \sigma_\mathrm{z}^{2}+\sigma_k^{2},
\end{align}
\end{subequations}
the optimization with respect to $\mathbf{W}$ can be simplified as
\begin{subequations}\label{pr:update_W}
\begin{align}
\max_{\mathbf{w}} ~~& \mathbf{w}^H\mathbf{Y}\mathbf{w}\\
\text{s.t.} ~~~& \mathbf{w}^H\mathbf{w} \leq P_{\mathrm{BS}},\\
& \mathbf{w}^H\mathbf{Z}\mathbf{w} \leq P_{\mathrm{RIS}} - c_\mathrm{RIS},\\
& \sqrt{1+\Gamma_{k}} |\mathbf{h}_k^{T}\mathbf{w}_k| \geq \sqrt{\Gamma_{k}} \big\| \big[{(\mathbf{H}}_{k}\mathbf{w})^T, \sqrt{c_0}\big]\big\|_2,~\forall k.
\end{align}
\end{subequations}
Since $\mathbf{Y}$ is a positive semi-definite Hermitian matrix according to its definition in (\ref{eq:trans_W}b), the objective function $\mathbf{w}^H\mathbf{Y}\mathbf{w}$ is convex. Obviously, maximizing a convex function makes the aforementioned problem non-convex. In order to deal with this challenging task, we exploit MM algorithm to convert it into a series of easily optimized problems until convergence. Specifically, with the obtained solution $\widetilde{\mathbf{w}}_{s}$ in the $s$-th iteration, we construct a more tractable surrogate function that approximates the objective function at the current point $\widetilde{\mathbf{w}}_{s}$ and serves as a lower-bound to be
maximized in the next iteration. Particularly, via the first-order Taylor expansion \cite{SunY}, an appropriate surrogate function for $\mathbf{w}^H\mathbf{Y}\mathbf{w}$ is given by
\begin{equation}\label{eq:W_first-order_Taylor_expansion}
  \mathbf{w}^H\mathbf{Y}\mathbf{w} \geq \widetilde{\mathbf{w}}_{s}^H\mathbf{Y}\widetilde{\mathbf{w}}_{s} + 2\Re\{\widetilde{\mathbf{w}}_{s}^H\mathbf{Y}(\mathbf{w}-\widetilde{\mathbf{w}}_{s})\}.
\end{equation}
Thus, the transmit beamformer design problem at point $\widetilde{\mathbf{w}}_{s}$ can be formulated as
\begin{equation}\label{pr:update_W_first-order}
\begin{aligned}
\max_{\mathbf{w}} ~~& \Re\{\widetilde{\mathbf{w}}_{s}^H\mathbf{Y}\mathbf{w}\}\\
\text{s.t.} ~~~& \text{(\ref{pr:update_W}b) - (\ref{pr:update_W}d)},
\end{aligned}
\end{equation}
which is a convex optimization problem and can be readily solved by various efficient algorithms.

\subsection{Reflection Coefficients Design}
When the receive filter $\mathbf{u}$ and the transmit beamforming $\mathbf{W}$ are obtained, the optimization problem solving for the reflection coefficients $\boldsymbol{\phi}$ is formulated as
\begin{equation}\label{pr:update_phi}
\begin{aligned}
\max_{\boldsymbol{\phi}} ~~&\gamma_\mathrm{r} = \frac{f(\boldsymbol{\phi})}{g(\boldsymbol{\phi})}\\
\text{s.t.} ~~~& \text{(\ref{pr:original_problem}c) - (\ref{pr:original_problem}e)},
\end{aligned}
\end{equation}
where $f(\boldsymbol{\phi})$ and $g(\boldsymbol{\phi})$ are notations presented as
\begin{subequations}\label{eq:fphi_gphi}
\begin{align}
f(\boldsymbol{\phi}) &\eqdef \varsigma_\mathrm{t}^2 \mathbf{u}^H\mathbf{H}_\mathrm{t}\mathbf{W}\mathbf{W}^H\mathbf{H}_\mathrm{t}^H\mathbf{u},\\
g(\boldsymbol{\phi}) &\eqdef \mathbf{u}^H(\varsigma_\mathrm{t}^2\sigma_\mathrm{z}^2\mathbf{H}_{\mathrm{z},0}\mathbf{H}_{\mathrm{z},0}^H + \sigma_\mathrm{z}^2\mathbf{H}_{\mathrm{z},1}\mathbf{H}_{\mathrm{z},1}^H + \sigma_\mathrm{r}^2\mathbf{I}_N)\mathbf{u}.
\end{align}
\end{subequations}
Unfortunately, this sub-problem is still difficult to directly solve due to the fractional objective and the implicit functions with respect to $\boldsymbol{\phi}$. To address these issues, we first re-formulate the fractional objective by applying Dinkelbach's transform and then recast the problem into an explicit form of $\boldsymbol{\phi}$.

Specifically, by introducing a new auxiliary variable $\varpi$, the transformed objective can be expressed as
\begin{equation}\label{pr:update_phi_FP}
\max_{\boldsymbol{\phi}} ~~ f(\boldsymbol{\phi}) - \varpi g(\boldsymbol{\phi}),
\end{equation}
where $\varpi$ has a closed-form solution in each iteration as
\begin{equation}\label{eq:varpi}
  \varpi^{\star} = \frac{f(\boldsymbol{\phi})}{g(\boldsymbol{\phi})}.
\end{equation}
Then, we attempt to extract variable $\boldsymbol{\phi}$ from the objective. By employing the transformation $\mathbf{\Phi}\mathbf{h}_{\mathrm{r,t}} = \text{diag}\{\mathbf{h}_{\mathrm{r,t}}\}\boldsymbol{\phi}$ and defining $\widetilde{\mathbf{G}} \eqdef \mathbf{G}^T\text{diag}\{\mathbf{h}_\mathrm{r,t}\}$, $\mathbf{H}_\mathrm{t}$ can be re-written as $\mathbf{H}_\mathrm{t} = \widetilde{\mathbf{G}}\boldsymbol{\phi}\boldsymbol{\phi}^T \widetilde{\mathbf{G}}^T$. Thus, $f(\boldsymbol{\phi})$ can be further expressed as
\begin{subequations}\label{eq:fphi}
\begin{align}
  f(\boldsymbol{\phi}) &= \varsigma_\mathrm{t}^2\mathbf{u}^H\widetilde{\mathbf{G}}\boldsymbol{\phi}\boldsymbol{\phi}^T \widetilde{\mathbf{G}}^T\mathbf{W}\mathbf{W}^H\widetilde{\mathbf{G}}^{*}{\boldsymbol{\phi}}^{*}{\boldsymbol{\phi}}^{H}\widetilde{\mathbf{G}}^{H}\mathbf{u}\\
  & = \varsigma_\mathrm{t}^2\text{Tr}\{\widetilde{\mathbf{G}}^{H}\mathbf{u}\mathbf{u}^H\widetilde{\mathbf{G}}\boldsymbol{\phi}\boldsymbol{\phi}^T \widetilde{\mathbf{G}}^T\mathbf{W}\mathbf{W}^H\widetilde{\mathbf{G}}^{*}{\boldsymbol{\phi}}^{*}{\boldsymbol{\phi}}^{H}\}\\
  & = \varsigma_\mathrm{t}^2\text{vec}^H\{\boldsymbol{\phi}\boldsymbol{\phi}^T\}\big[(\widetilde{\mathbf{G}}^H\mathbf{W}^{*}\mathbf{W}^{T}\widetilde{\mathbf{G}}) \otimes (\widetilde{\mathbf{G}}^{H}\mathbf{u}\mathbf{u}^H\widetilde{\mathbf{G}})\big]
  \notag\\
  & \quad ~ \text{vec}\{\boldsymbol{\phi}\boldsymbol{\phi}^T\}\\
  & = \mathbf{x}^H \mathbf{C}\mathbf{x},
\end{align}
\end{subequations}
where the equation (\ref{eq:fphi}c) holds by applying the transformation $\text{Tr}\{\mathbf{A}\mathbf{B}\mathbf{C}\mathbf{D}\} = \text{vec}^H\{\mathbf{D}^H\}(\mathbf{C}^T \otimes \mathbf{A})\text{vec}\{\mathbf{B}\}$,
and in (\ref{eq:fphi}d) we define $\mathbf{x} \eqdef \text{vec}\{\boldsymbol{\phi}\boldsymbol{\phi}^T\} = \boldsymbol{\phi} \otimes \boldsymbol{\phi}$ and
$\mathbf{C}  \eqdef \varsigma_\mathrm{t}^2(\widetilde{\mathbf{G}}^H\mathbf{W}^{*}\mathbf{W}^{T}\widetilde{\mathbf{G}}) \otimes (\widetilde{\mathbf{G}}^{H}\mathbf{u}\mathbf{u}^H\widetilde{\mathbf{G}})$.
Using the similar derivations in (\ref{eq:fphi}), we can equivalently re-formulate $g(\boldsymbol{\phi})$ as
\begin{equation}\label{eq:gphi}
\begin{aligned}
  g(\boldsymbol{\phi}) = \mathbf{x}^H\mathbf{D}\mathbf{x} + \boldsymbol{\phi}^H\mathbf{E}\boldsymbol{\phi}+ \sigma_\mathrm{r}^2\|\mathbf{u}\|_2^2,
\end{aligned}
\end{equation}
where we define $\mathbf{D} \eqdef \varsigma_\mathrm{t}^2\sigma_\mathrm{z}^2 (\text{diag}\{\mathbf{h}_\mathrm{r,t}\}\text{diag}\{\mathbf{h}_\mathrm{r,t}^*\}) \otimes (\widetilde{\mathbf{G}}^H\mathbf{u}\mathbf{u}^H\widetilde{\mathbf{G}})$ and $\mathbf{E} \eqdef \sigma_\mathrm{z}^2\text{diag}\{\mathbf{G}^*\mathbf{u}\}\text{diag}\{\mathbf{G}\mathbf{u}^*\}$.
According to (\ref{eq:fphi}) and (\ref{eq:gphi}), the objective in (\ref{pr:update_phi_FP}) can be re-arranged as
\begin{equation}\label{eq:update_phi_FP_fun}
\begin{aligned}
\min_{\boldsymbol{\phi}} ~~ \varpi g(\boldsymbol{\phi}) - f(\boldsymbol{\phi})
 = \mathbf{x}^H \mathbf{F} \mathbf{x} + \varpi\boldsymbol{\phi}^H \mathbf{E}\boldsymbol{\phi} + c_1,
\end{aligned}
\end{equation}
by defining the affine relationship $\mathbf{F} \eqdef  \varpi \mathbf{D} - \mathbf{C}$ and $c_1 \eqdef \varpi\sigma_\mathrm{r}^2\|\mathbf{u}\|_2^2$.

However, it is clear that $\mathbf{x}^H \mathbf{F} \mathbf{x}$ is a quartic term with respect to $\boldsymbol{\phi}$, which leads to an intractable objective. Therefore, we employ the idea of the MM method again to find a favorable surrogate function for it. In particular, via the second-order Taylor expansion, an appropriate upper-bound for $\mathbf{x}^H \mathbf{F}\mathbf{x}$ at point $\mathbf{x}_s$ (i.e., $\boldsymbol{\phi}_s$) can be derived as
\begin{equation}\label{eq:phi_second-order_Taylor_expansion_objective_1}
\begin{aligned}
  \mathbf{x}^H \mathbf{F}\mathbf{x} &\leq \lambda_\mathrm{f} \mathbf{x}^H\mathbf{x} + 2\Re\{\mathbf{x}^H(\mathbf{F} - \lambda_\mathrm{f}\mathbf{I}_{M^2})\mathbf{x}_{s}\}\\
     & \quad + \mathbf{x}_s^H(\lambda_\mathrm{f}\mathbf{I}_{M^2} - \mathbf{F})\mathbf{x}_{s},
\end{aligned}
\end{equation}
where $\lambda_\mathrm{f}$ is the maximum eigenvalue of matrix $\mathbf{F}$. Taking the amplitude constraint $a_m \leq a_{\max}$ into consideration, it is clear that the term $\mathbf{x}^H\mathbf{x}$ is upper-bounded by
\begin{equation}\label{eq:MM_x_phi}
\mathbf{x}^H\mathbf{x} \hspace{-0.05cm}=\hspace{-0.05cm} (\boldsymbol{\phi} \otimes  \boldsymbol{\phi})^H\hspace{-0.05cm}(\boldsymbol{\phi} \otimes \boldsymbol{\phi})\hspace{-0.05cm} =\hspace{-0.05cm} (\boldsymbol{\phi}^H\hspace{-0.05cm}\boldsymbol{\phi})\otimes(\boldsymbol{\phi}^H\hspace{-0.05cm}\boldsymbol{\phi}) \hspace{-0.05cm}\leq\hspace{-0.05cm} M^2\alpha_{\max}^4.
\end{equation}
Substituting the result in (\ref{eq:MM_x_phi}) into (\ref{eq:phi_second-order_Taylor_expansion_objective_1}), we further obtain
\begin{subequations}\label{eq:phi_second-order_Taylor_expansion_objective_1_2}
\begin{align}
   \mathbf{x}^H \mathbf{F}\mathbf{x}  &\leq   \Re\{\mathbf{x}^H \mathbf{f}\} + c_2,\\
    & = \Re\{\boldsymbol{\phi}^H\widetilde{\mathbf{F}} \boldsymbol{\phi}^{*}\} + c_2,
\end{align}
\end{subequations}
where we define the scalar $c_2 \eqdef \lambda_\mathrm{f} {M^2} \alpha_\mathrm{\max}^4 + \mathbf{x}_s^H(\lambda_\mathrm{f}\mathbf{I}_{M^2} - \mathbf{F})\mathbf{x}_{s}$ irrelevant to the variable $\mathbf{x}$, $\mathbf{f} \triangleq 2(\mathbf{F} - \lambda_\mathrm{f}\mathbf{I}_{M^2})\mathbf{x}_{s}$, and $\widetilde{\mathbf{F}}$ is the matrix form of vector $\mathbf{f}$, i.e., $\mathbf{f} \eqdef \text{vec}\{\widetilde{\mathbf{F}}\}$.
Since $\Re\{\boldsymbol{\phi}^H\widetilde{\mathbf{F}}\boldsymbol{\phi}^*\}$ in (\ref{eq:phi_second-order_Taylor_expansion_objective_1_2}b) is still non-convex with respect to $\boldsymbol{\phi}$, we propose to equivalently convert the complex-valued function into its real-valued form $\bar{\boldsymbol{\phi}}^T \bar{\mathbf{F}}\bar{\boldsymbol{\phi}}$ by defining $\bar{\boldsymbol{\phi}} \triangleq \begin{bmatrix}\Re\{\boldsymbol{\phi}^{T}\} ~~ \Im\{\boldsymbol{\phi}^{T}\}\end{bmatrix}^{T}$ and $\bar{\mathbf{F}} \triangleq \begin{bmatrix}
\Re\{\widetilde{\mathbf{F}}\} & \Im\{\widetilde{\mathbf{F}}\} \\
\Im\{\widetilde{\mathbf{F}}\} & -\Re\{\widetilde{\mathbf{F}}\}
\end{bmatrix}$.
Afterwards, we derive a convex surrogate function for $\bar{\boldsymbol{\phi}}^T \bar{\mathbf{F}}\bar{\boldsymbol{\phi}}$ by the second-Taylor expansion again as:
\begin{subequations}\label{eq:phi_second-order_Taylor_expansion_objective_2}
\begin{align}
  \hspace{-0.5cm}  \bar{\boldsymbol{\phi}}^T \bar{\mathbf{F}}\bar{\boldsymbol{\phi}}
  & \leq \bar{\boldsymbol{\phi}}_s^T \bar{\mathbf{F}} \bar{\boldsymbol{\phi}}_s + \bar{\boldsymbol{\phi}}_s^T (\bar{\mathbf{F}} + \bar{\mathbf{F}}^T)(\bar{\boldsymbol{\phi}} - \bar{\boldsymbol{\phi}}_s)\notag\\
   &\quad+ \frac{\lambda_\mathrm{\tilde{f}}}{2}(\bar{\boldsymbol{\phi}} - \bar{\boldsymbol{\phi}}_s)^T(\bar{\boldsymbol{\phi}} - \bar{\boldsymbol{\phi}}_s)\\
  & = \frac{\lambda_\mathrm{\tilde{f}}}{2} \boldsymbol{\phi}^H\boldsymbol{\phi} + \Re\{ \boldsymbol{\phi}^H\widetilde{\mathbf{f}}\} + c_3,
\end{align}
\end{subequations}
where $\lambda_\mathrm{\tilde{f}}$ is the maximum eigenvalue of Hessian matrix $(\bar{\mathbf{F}} + \bar{\mathbf{F}}^T)$, $\widetilde{\mathbf{f}} \eqdef \mathbf{U} (\bar{\mathbf{F}} + \bar{\mathbf{F}}^T - \lambda_\mathrm{\tilde{f}}\mathbf{I}_{2M})\bar{\boldsymbol{\phi}}_s$, $\mathbf{U} \eqdef \left[\mathbf{I}_{M} ~ \jmath\mathbf{I}_{M}\right]$,
and $c_3 \eqdef -\bar{\boldsymbol{\phi}}_s^T\bar{\mathbf{F}}^T\bar{\boldsymbol{\phi}}_s + \frac{\lambda_\mathrm{\tilde{f}}}{2}\bar{\boldsymbol{\phi}}_s^T\bar{\boldsymbol{\phi}}_s$ is a scalar independent of ${\boldsymbol{\phi}}$. After summarizing the derivations in (\ref{eq:phi_second-order_Taylor_expansion_objective_1}), (\ref{eq:phi_second-order_Taylor_expansion_objective_1_2}) and (\ref{eq:phi_second-order_Taylor_expansion_objective_2}), we have
\begin{equation}\label{eq:final_xFx}
 \mathbf{x}^H \mathbf{F}\mathbf{x} \leq \frac{\lambda_\mathrm{\tilde{f}}}{2} \boldsymbol{\phi}^H\boldsymbol{\phi} + \Re\{ \boldsymbol{\phi}^H\widetilde{\mathbf{f}}\} + c_3 + c_2.
\end{equation}
Therefore, the objective function of optimizing $\boldsymbol{\phi}$ in (\ref{eq:update_phi_FP_fun}) can be converted into
\vspace{-0.2cm}
\begin{equation}\label{eq:final_objective}
\min_{\boldsymbol{\phi}} ~~ \boldsymbol{\phi}^H \widetilde{\mathbf{E}}\boldsymbol{\phi} + \Re\{\boldsymbol{\phi}^H\widetilde{\mathbf{f}}\}
\end{equation}
with $\widetilde{\mathbf{E}} \eqdef \varpi\mathbf{E} + \frac{\lambda_\mathrm{\tilde{f}}}{2} \mathbf{I}_M$.

Nevertheless, the transformed problem is also non-convex due to the non-convex constraint in (\ref{pr:original_problem}c). In order to
facilitate the algorithm development, we first extract the variable $\boldsymbol{\phi}$ from $\mathcal{P}(\boldsymbol{\phi})$ and equivalently re-write it as
\begin{equation}\label{eq:P_phi}
\begin{aligned}
  \mathcal{P}(\boldsymbol{\phi}) = \mathbf{x}^H\mathbf{J}\mathbf{x} + \boldsymbol{\phi}^H\mathbf{K}\boldsymbol{\phi}
\end{aligned}
\end{equation}
by defining the following transformations:
\begin{subequations}\label{eq:P_phi_trans}
\begin{align}
  \mathbf{J} &\eqdef \varsigma_\mathrm{t}^2 (\widetilde{\mathbf{G}}^T\mathbf{W}\mathbf{W}^H\widetilde{\mathbf{G}}^* )^T \otimes \widetilde{\mathbf{J}} + \varsigma_\mathrm{t}^2 \sigma_\mathrm{z}^2 \widetilde{\mathbf{J}} \otimes \widetilde{\mathbf{J}},\\
   \widetilde{\mathbf{J}} &\eqdef \text{diag}\{\mathbf{h}_\mathrm{r,t}\}\text{diag}\{\mathbf{h}_\mathrm{r,t}^*\},\\
  \mathbf{K} &\eqdef \sum\nolimits_{k = 1}^{K+N} \text{diag}\{\mathbf{G}^*\mathbf{w}_k^*\} \text{diag}\{\mathbf{G}\mathbf{w}_k\} + 2\sigma_\mathrm{z}^2 \mathbf{I}_{M}.
\end{align}
\end{subequations}
By utilizing the same MM-based algorithmic framework in (\ref{eq:phi_second-order_Taylor_expansion_objective_1})-(\ref{eq:final_xFx}), a convex surrogate function for the quartic term $\mathbf{x}^H\mathbf{J}\mathbf{x}$ in (\ref{eq:P_phi}) can be constructed  as
\begin{subequations}\label{eq:p_phi_MM1_3}
\begin{align}
    \mathbf{x}^H \mathbf{J}\mathbf{x} &\leq \Re\{\boldsymbol{\phi}^H\widetilde{\mathbf{P}} \boldsymbol{\phi}^{*}\} + c_4 = \bar{\boldsymbol{\phi}}^T \bar{\mathbf{P}} \bar{\boldsymbol{\phi}} + c_4\\
    &\leq \frac{\lambda_\mathrm{p}}{2} \boldsymbol{\phi}^H\boldsymbol{\phi} + \Re\{\boldsymbol{\phi}^H \widetilde{\mathbf{p}}\} + c_5,
\end{align}
\end{subequations}
where $\widetilde{\mathbf{P}}$ is the matrix form of vector $\mathbf{p} \triangleq 2(\mathbf{J} - \lambda_\mathrm{j}\mathbf{I}_{M^2})\mathbf{x}_{s}$, $\bar{\mathbf{P}} \triangleq \begin{bmatrix}
\Re\{\widetilde{\mathbf{P}}\} & \Im\{\widetilde{\mathbf{P}}\} \\
\Im\{\widetilde{\mathbf{P}}\} & -\Re\{\widetilde{\mathbf{P}}\}
\end{bmatrix}$ is a real-valued matrix corresponding to $\widetilde{\mathbf{P}}$ and $c_4 \eqdef \lambda_\mathrm{j} {M^2} \alpha_\mathrm{\max}^4 + \mathbf{x}_s^H(\lambda_\mathrm{j}\mathbf{I}_{M^2} - \mathbf{J})\mathbf{x}_{s}$ with $\lambda_\mathrm{j} \eqdef \text{Tr}\{\mathbf{J}\}$. In addition, we define $\lambda_\mathrm{p}$ as the maximum eigenvalue of Hessian matrix $(\bar{\mathbf{P}} + \bar{\mathbf{P}}^T)$, $\widetilde{\mathbf{p}} \eqdef \mathbf{U} (\bar{\mathbf{P}} + \bar{\mathbf{P}}^T - \lambda_\mathrm{p}\mathbf{I}_{2M})\bar{\boldsymbol{\phi}}_s$ and $c_5 \eqdef  -\bar{\boldsymbol{\phi}}_s^T\bar{\mathbf{P}}^T\bar{\boldsymbol{\phi}}_s + \frac{\lambda_\mathrm{p}}{2}\bar{\boldsymbol{\phi}}_s^T\bar{\boldsymbol{\phi}}_s + c_4$.

With the surrogate function of $\mathbf{x}^H \mathbf{J}\mathbf{x}$ derived in (\ref{eq:p_phi_MM1_3}),
a tractable upper-bound function for reflection power $\mathcal{P}(\boldsymbol{\phi})$ can be provided by
\begin{equation}\label{eq:p_phi_MM1_4}
   \mathcal{P}(\boldsymbol{\phi})\leq \boldsymbol{\phi}^H \widetilde{\mathbf{K}} \boldsymbol{\phi} + \Re\{\boldsymbol{\phi}^H \widetilde{\mathbf{p}}\} + c_5,
\end{equation}
where we define $\widetilde{\mathbf{K}} \eqdef \mathbf{K} +\frac{\lambda_\mathrm{p}}{2}\mathbf{I}_M$ for simplicity.

Consequently, the optimization for updating $\boldsymbol{\phi}$ can be re-formulated as
\begin{subequations}\label{pr:final_phi}
\begin{align}
\min_{\boldsymbol{\phi}} ~~& \boldsymbol{\phi}^H \widetilde{\mathbf{E}}\boldsymbol{\phi} + \Re\{\boldsymbol{\phi}^H\widetilde{\mathbf{f}}\}\\
\text{s.t.} ~~~
& \boldsymbol{\phi}^H \widetilde{\mathbf{K}} \boldsymbol{\phi} + \Re\{\boldsymbol{\phi}^H \widetilde{\mathbf{p}}\}  \leq P_\mathrm{RIS} - c_5,\\
& \sqrt{1+{\Gamma_k}} |\widetilde{a}_{k}(\boldsymbol{\phi})| \geq \sqrt{{\Gamma_k}} \big\| \widetilde{\mathbf{b}}_{k}(\boldsymbol{\phi})\big\|_2,~\forall k,\\
& a_m \leq a_{\max},~\forall m,
\end{align}
\end{subequations}
where $\widetilde{a}_{k}(\boldsymbol{\phi})$ and $\widetilde{\mathbf{b}}_{k}(\boldsymbol{\phi})$ are both linear functions with respect to $\boldsymbol{\phi}$, which can be expressed as
\begin{subequations}\label{eq:final_phi_trans}
\begin{align}
\widetilde{a}_{k}(\boldsymbol{\phi}) &\eqdef \mathbf{h}_{\mathrm{d},k}^{T}\mathbf{w}_k + \mathbf{h}_{\mathrm{r},k}^T\text{diag}\{\mathbf{G}\mathbf{w}_k\}\boldsymbol{\phi},\\
\widetilde{\mathbf{b}}_{k}(\boldsymbol{\phi}) &\eqdef \big[(\mathbf{a}_k + \mathbf{B}_k^T \boldsymbol{\phi})^T, (\sigma_\mathrm{z}\text{diag}\{\mathbf{h}_{\mathrm{r},k} \} \boldsymbol{\phi})^T,\sigma_k\big]^T,\\
\mathbf{a}_k &\eqdef [\mathbf{h}_{\mathrm{d},k}^{T}\mathbf{w}_1, \cdots, \mathbf{h}_{\mathrm{d},k}^{T}\mathbf{w}_{K+N}]^T,\\
\mathbf{B}_k &\eqdef [\text{diag}\{\mathbf{G}\mathbf{w}_1\}\mathbf{h}_{\mathrm{r},k}, \cdots, \text{diag}\{\mathbf{G}\mathbf{w}_{K+N}\}\mathbf{h}_{\mathrm{r},k}].
\end{align}
\end{subequations}
Now, problem (\ref{pr:final_phi}) is convex that can be solved optimally by existing optimization algorithms or solvers, e.g., CVX.

\subsection{Summary}
Based on the above derivations, the joint transceiver beamforming and active RIS reflection design for the considered active RIS-aided ISAC system is straightforward. In each iteration, the receiver filter $\mathbf{u}$, the transmit beamformer $\mathbf{W}$ and the active RIS reflection coefficients $\boldsymbol{\phi}$ are iteratively updated until convergence. The total computational complexity of the proposed algorithm is approximated at the order of $\mathcal{O}([N(K+N)]^{4.5} + M^{4.5})$.

\section{Simulation Results}

\begin{figure*}[t]
  \begin{minipage}[t]{0.31\textwidth}
    \centering
    \includegraphics[width=2.4 in]{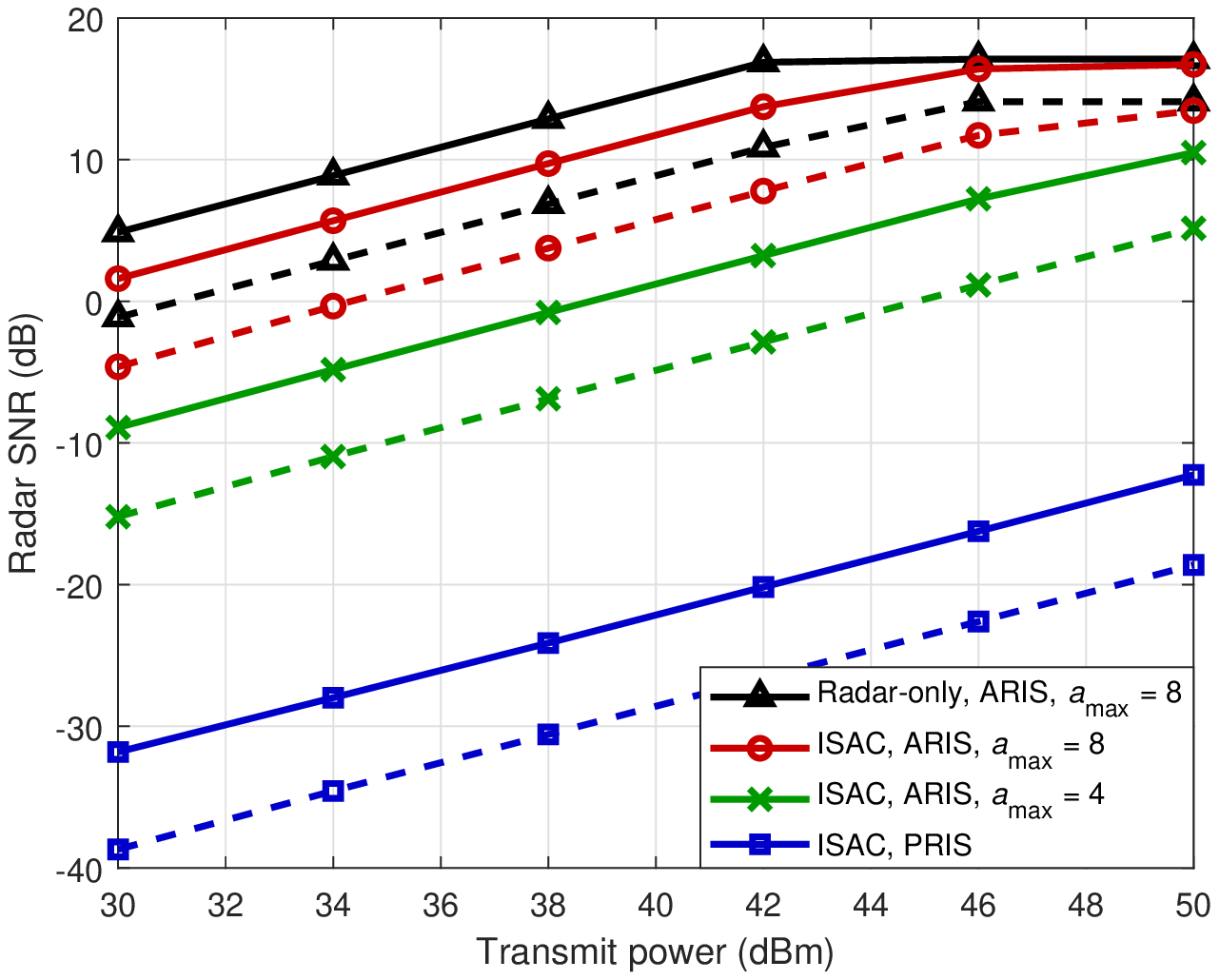}
    \caption{Radar SNR $\gamma_\mathrm{r}$ versus the transmit power $P_\mathrm{BS}$ ($M = 32$, $\Gamma = 12$dB, solid: $N = 16$, dash: $N = 8$).}
    \label{fig:Pb}
  \end{minipage}%
  \hspace{0.3cm}
  \begin{minipage}[t]{0.31\textwidth}
    \centering
    \includegraphics[width =2.4 in]{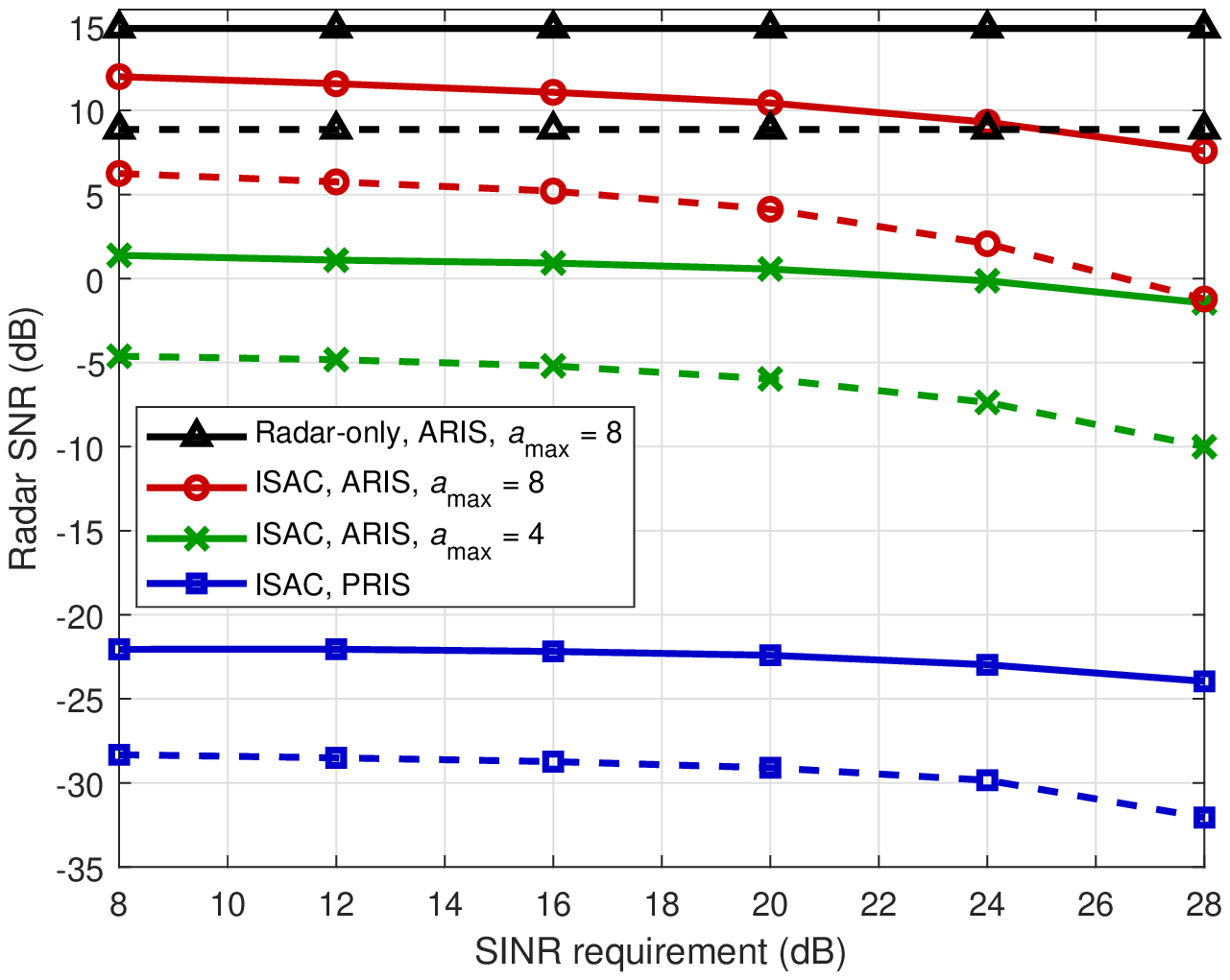}
    \caption{Radar SNR $\gamma_\mathrm{r}$ versus the SINR requirement $\Gamma$ ($P_\mathrm{BS}$ = 40dBm, $M = 32$, solid: $N = 16$, dash: $N = 8$).}
    \label{fig:sinr}
  \end{minipage}%
  \hspace{0.3cm}
  \begin{minipage}[t]{0.33\textwidth}
    \centering
    \includegraphics[width =2.4 in]{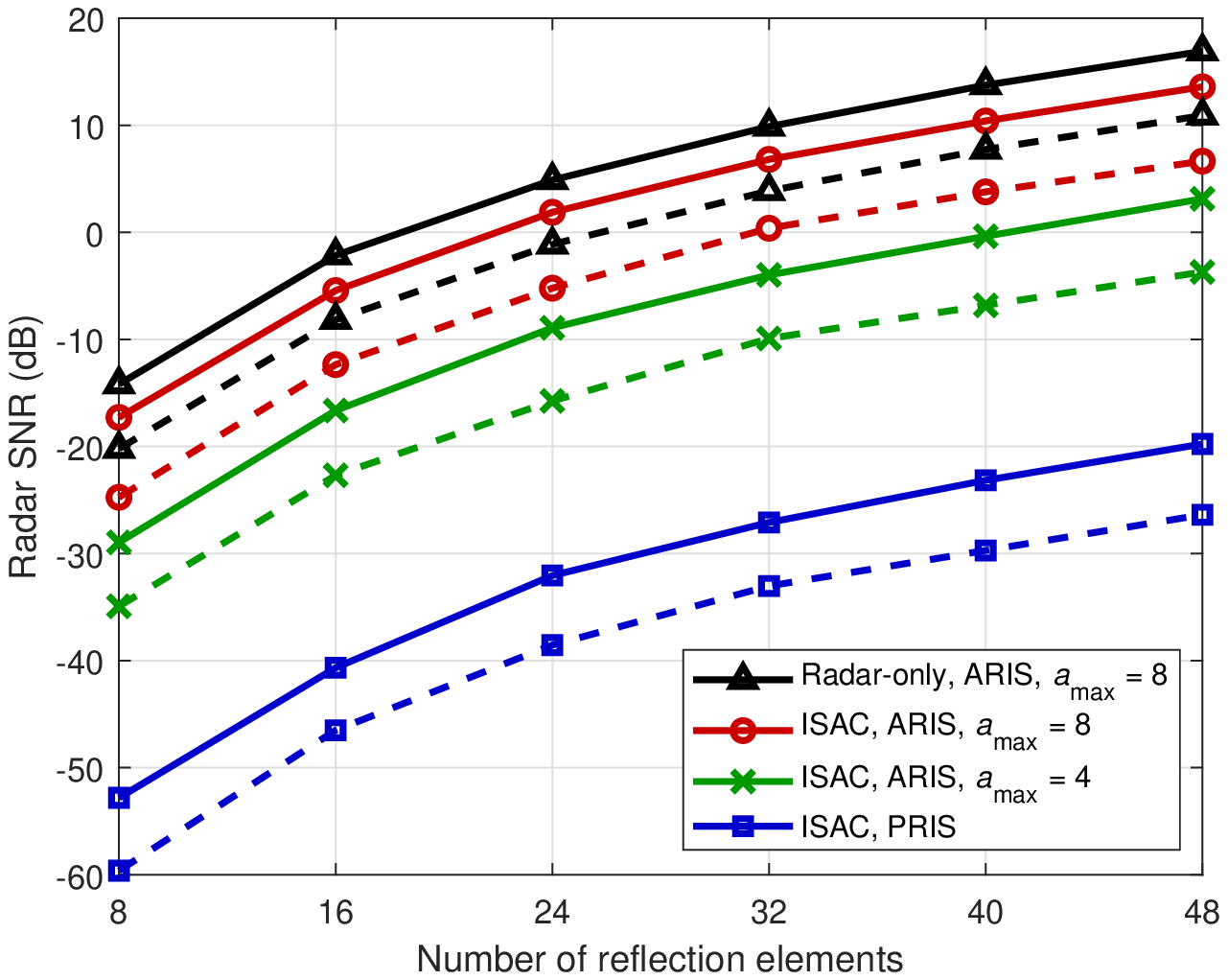}
    \caption{Radar SNR $\gamma_\mathrm{r}$ versus the number of RIS elements $M$ ($P_\mathrm{BS}$ = 35dBm, $\Gamma = 12$dB, solid: $N = 16$, dash: $N = 8$).}
    \label{fig:M}
  \end{minipage}%
\vspace{-0.3cm}
\end{figure*}

This section provides simulation results to demonstrate the advancement of proposed active RIS-assisted ISAC scheme and the effectiveness of developed joint design algorithm. We set the numbers of users and elements of active RIS to $K = 4$ and $M = 32$, respectively.
For simplicity, we assume the SINR requirements are the same for all users, i.e., $\Gamma_k = \Gamma$. The power budget at the active RIS is set as $P_\mathrm{RIS} = 20$dBm. Furthermore, we set the noise powers as $\sigma_k^2 = \sigma_\mathrm{z}^2 = \sigma_\mathrm{r}^2 = -80$dBm,$~\forall k$, and the RCS as $\varsigma_\mathrm{t}^2 = 1$.
Moreover, the channel models and parameters are configured similar to \cite{LiuR1}.

We first present the radar output SNR $\gamma_\mathrm{r}$ versus the transmit power $P_\mathrm{BS}$ in Fig. \ref{fig:Pb}. In addition to the proposed active RIS-assisted ISAC scheme (``ISAC, ARIS"), we also include the passive RIS-assisted ISAC scheme (``ISAC, PRIS") and the active RIS-assisted radar-only scheme (``Radar-only, ARIS") as comparisons. For fairness, we set $P_\mathrm{BS} + P_\mathrm{RIS}$ as the transmit power for the passive RIS case. It is observed from Fig. \ref{fig:Pb} that at the case of $a_{\max} = 8$, the active RIS-assisted ISAC scheme can achieve up to 32dB radar performance improvement compared to the scheme with passive RIS when $P_\mathrm{BS} \leq 42$dBm, which verifies the superiority of deploying active RIS in ISAC systems when the potential target is located in the dead zone of the BS.
Moreover, since high-quality communication performance is guaranteed, the active RIS-aided ISAC scheme has about 3dB radar performance loss compared with the radar-only scheme when $P_\mathrm{BS} \leq 42$dBm.
With the growth of $P_\mathrm{BS}$, the radar SNR of systems with the aid of active RIS at $a_{\max} = 8$ case first increases rapidly, and then tends to be flat or even constant. The reason for this phenomenon is that the fixed power budget $P_\mathrm{RIS}$ at the active RIS limits the radar performance even if transmit power $P_\mathrm{BS}$ grows.
In addition, even if $a_{\max}$ is reduced to 4, the active RIS-assisted ISAC scheme is still significantly superior to the passive RIS scheme.


The radar output SNR $\gamma_\mathrm{r}$ versus the users' SINR requirement $\Gamma$ is illustrated in Fig. \ref{fig:sinr}. We can notice that the increase of communication quality requirement leads to the degradation of radar sensing performance, which reveals the trade-off between the radar and communication performance. However, the phenomenon is not obvious when $\Gamma$ is smaller, i.e., 8dB-16dB. It is because that when maximizing radar SNR, the resulting directional beams towards active RIS can readily satisfy the communication requirements.

Fig. \ref{fig:M} depicts the radar output $\gamma_\mathrm{r}$ versus the number of reflection elements $M$. Not surprisingly, the radar SNR increases with the growth in the number of reflecting elements since it provides more DoFs to manipulate wireless environment. Moreover, the active RIS-assisted ISAC scheme can always dramatically outperform the passive RIS scheme under different $M$.

\section{Conclusions}
In this paper, we investigated the joint design of transmit beamformer, the active RIS reflection and the radar receive filter for active RIS-aided ISAC systems. An efficient algorithm utilizing BCD, Dinkelbach's transform and MM methods was developed to maximize the radar output SNR while ensuring the SINR performance of communication users. Simulation studies demonstrated the significant advantages of deploying active RIS in ISAC systems, as well as the effectiveness of proposed design algorithm.

\end{document}